%% file: beuther.tex
\documentclass{kapproc} 
\usepackage{boxedminipage}
\RequirePackage{graphicx}%
\RequirePackage{epsf}%
\input{psfig.sty}

\upperandlowercase
\let\footnote\savefootnote
\let\footnotetext\savefootnotetext 
\let\footnoterule\savefootnoterule 
\kluwerbib 

\begin{document}
\input{beuther_refs.tex}


\articletitle{Fragmentation of a high-mass star-forming core}


\chaptitlerunninghead{Fragmentation in massive star formation}



 \author{H. Beuther}
 \affil{Harvard-Smithsonian Center for Astrophysics, 60 Garden Street, Cambridge MA 02138, USA}
 \email{hbeuther@cfa.harvard.edu}





 \begin{abstract}
At the earliest evolutionary stages, massive star-forming regions are
deeply embedded within their natal cores and not observable at optical
and near-infrared wavelengths. Interferometric high-spatial resolution
mm dust continuum observations of one very young high-mass
star-forming region disentangle its cluster-like nature already at the
very beginning of the star formation process. The derived
protocluster mass function is consistent with the stellar IMF. Hence,
fragmentation of the initial massive cores may determine the IMF and
the masses of the final stars. This implies that stars of all masses
can form via accretion, and coalescence of protostars appears not to
be necessary.
 \end{abstract}


\noindent {\bf Introduction}\\ One mystery in star formation is at
what evolutionary stage in the cluster formation process the shape of
the IMF gets established. For low-mass clusters, \cite{motte1998} have
shown that the protocluster mass function of $\rho$ Ophiuchus
resembles already the final IMF, but so far no comparable study exist
for very young high-mass star-forming regions. Additionally, I like to
address how far studying the earliest fragmentation processes of
massive star-forming regions helps to differentiate between the two
proposed scenarios for massive star formation: accretion versus
coalescence (e.g., \cite{mckee2002,bonnell1998}). Employing the
Plateau de Bure Interferometer (PdBI), we studied the dust continuum
emission at 1.3 and 3\,mm toward the massive star-forming region
IRAS\,19410+2336. This region of $10^4$\,L$_{\odot}$ at a distance of
$\sim 2$\,kpc is at an early evolutionary stage prior to forming a hot
core. The results of this study have recently been reported by
\cite{beuther2004c}.\\

\begin{figure}[htb] \begin{center}
\includegraphics[angle=-90,width=10cm]{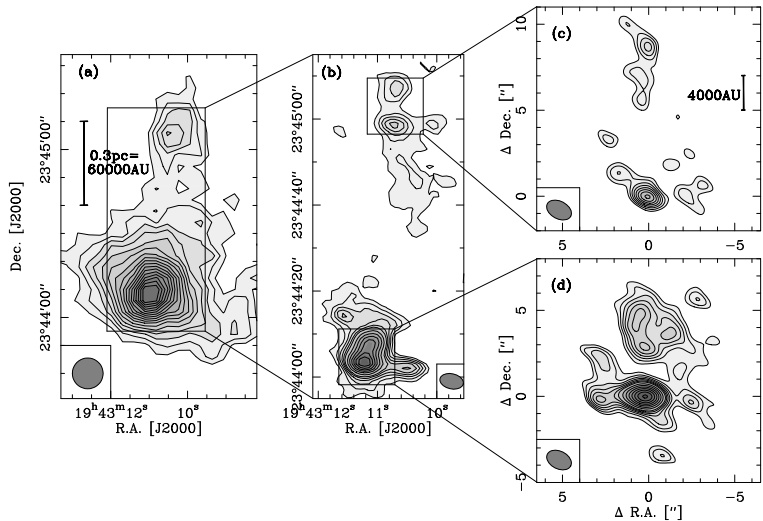} \end{center}
\caption{Dust continuum images of IRAS\,19410+2336. Left: 1.2\,mm
  single-dish data obtained with the IRAM 30\,m at $11''$
  (\cite{beuther2002a}). Middle and right: 3\,mm and 1.3\,mm PdBI data
  obtained with spatial resolutions nearly an order of magnitude
  better ($5.5''\times 3.5''$ and $1.5''\times1''$, respectively).}
\label{fig1}
\end{figure}

\noindent {\bf Results}\\ The large scale continuum emission observed
at 1.2\,mm with the IRAM 30\,m telescope (Fig.\,\ref{fig1}a) shows two
massive gas cores. Based on the intensity profiles, we predicted that
the cores should split up into sub-structures at scales between $3''$
and $5''$ (\cite{beuther2002a}). The PdBI 3\,mm data at more than
twice the spatial resolution show that both sources split up into
sub-structures at the predicted scales, about four sources in the
southern core and four in the northern core (Fig.\,\ref{fig1}b).  At
the highest spatial resolution (Figs.\,\ref{fig1}c \& d), we observe
that the gas clumps resolve into even more sub-sources. We find small
clusters of gas and dust condensations with 12 sources per
protocluster over the $3\sigma$ limit of 9\,mJy/beam. Both
protoclusters are dominated by one central massive source and
surrounded by a cluster of less massive sources. This provides
evidence for the fragmentation of a high-mass protocluster down to
scales of 2000\,AU at the earliest evolutionary stages.

Assuming the mm continuum flux to be due to optically thin thermal
dust emission, one can calculate the masses following the method
outlined in \cite{hildebrand1983}. Based on IRAS far-infrared
observations we estimate the average dust temperature to be around
46\,K (\cite{sridha}), the dust opacity index $\beta$ is set to 2
(\cite{beuther2002a}). At the given temperature, the 9\,mJy/beam
sensitivity corresponds to a mass sensitivity limit of
1.7\,M$_{\odot}$. The range of clump masses is 1.7 to
25\,M$_{\odot}$. Combining the data from both clusters, we can derive
a protocluster mass spectrum $\Delta N/\Delta M$, with the number of
clumps $\Delta N$ per mass bin $\Delta M$ (Fig.\,2). The best fit to
the data results in a mass spectrum $\Delta N/\Delta M \propto M^{-a}$
with the power law index $a=2.5$ and a mean deviation $da=0.3$.

\vspace{0.3cm}
\begin{minipage}{6cm}
\includegraphics[angle=-90,width=6cm]{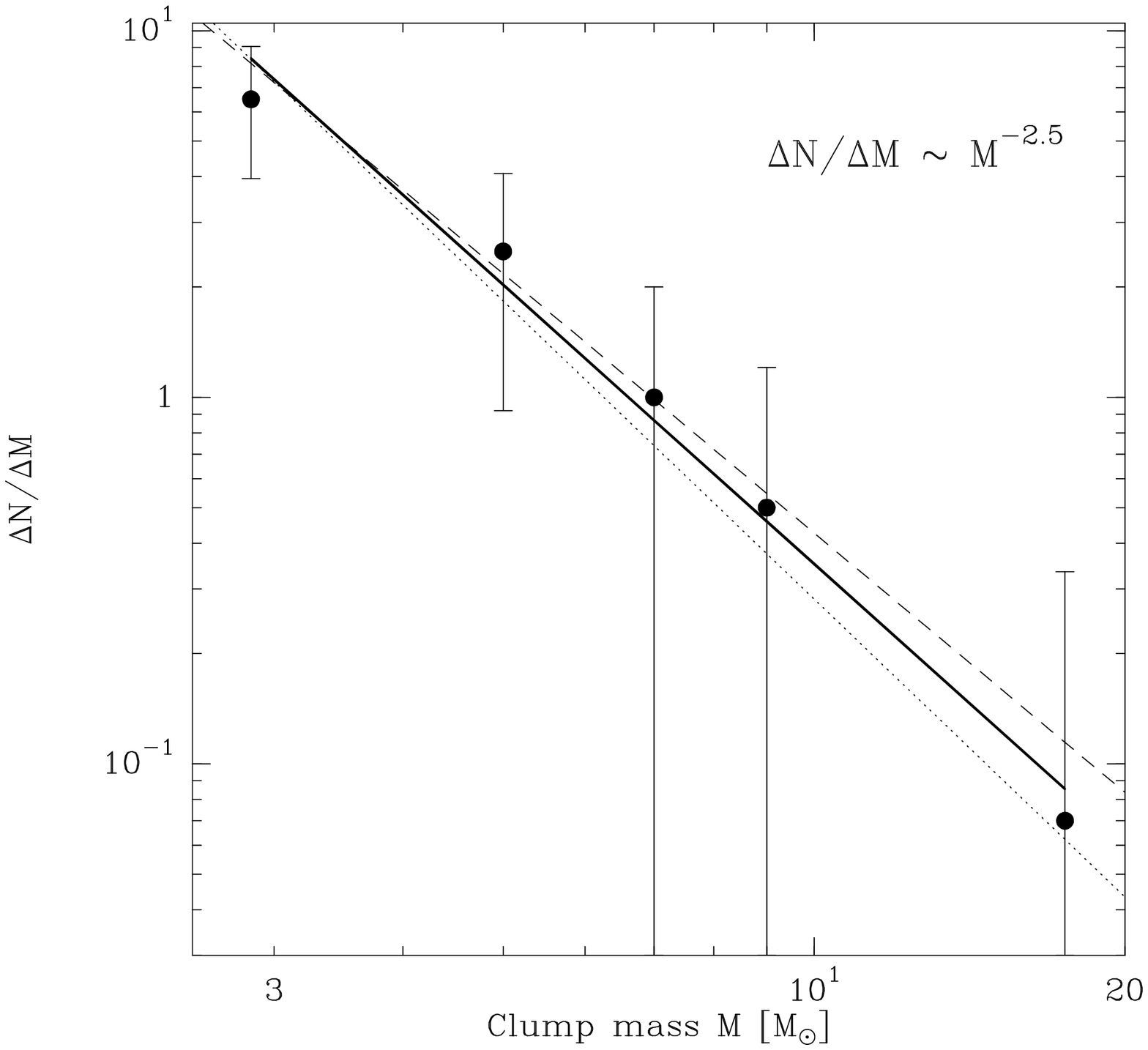}
\end{minipage}
\begin{minipage}{5cm}
\vspace{-0.4cm} Fig. 2: {\footnotesize The mass spectrum of
IRAS\,19410+2336.  The clump-mass bins are [1.7($3\sigma$),4], [4,6],
[6,8], [8,10] and [10,25] \,M$_{\odot}$. The best fit to the data
$\Delta N/\Delta M \sim M^{-a}$ is $a=2.5$, the dashed and dotted
lines show the IMFs derived from Salpeter (1955) with $a=2.35$ and
Kroupa et al.\,(1993) with $a=2.7$, respectively.}
\end{minipage}
\vspace{0.3cm}

{\it Caveats:} One uncertainty is the assumption of a uniform dust
temperature because higher temperatures for the massive clumps would
decrease their masses whereas lower temperatures for the less massive
clumps would increase those. These effects could result in a somewhat
flattened slope $a$. However, the protocluster is at an early
evolutionary stage prior to forming a significant hot core, and the
temperature variations in the protocluster should still be
small. Hence, it is plausible to assume a similar dust temperature for
all sub-sources, and the relative accuracy between the clumps
masses~-- and thus the slope of the mass spectrum~-- is high. An
additional caveat is that we are dealing with low-number statistics
and just five mass bins. Only future more sensitive observations of a
statistical significant number of massive protoclusters will allow to
handle the statistics better. In contrast to this, we do not believe
that the spatial filtering properties of the interferometer affect our
results because the size of all clumps is far smaller than the spatial
structures filtered out (of sizes $>20''$). Consequently, only a
large-scale halo common to all sources is affected by the filtering,
whereas the sources we are interested in are not. Another uncertainty
is whether all mm continuum emission is really due to protostellar
condensations, because \cite{gueth2003} have shown that such emission
can also be caused by molecular outflows. Since at least seven
outflows are observed toward IRAS\,19410+2336 (\cite{beuther2003a}),
one or the other emission feature could be due an outflow as
well. However, outflow associated emission features are expected to be
rather weak. As a consistency check, we increased the lower mass limit
slightly including less clumps in the power spectrum. The resulting
values of the power-law index $a$ varied only within the error
margins. Therefore, we conclude that possible outflow contributions should
not alter the derived slope $a$ significantly.\\

\noindent {\bf Discussion}\\ The derived power-law spectrum can be
compared with the IMFs of more evolved clusters. Although open issues
remain, this conference has confirmed the general consensus that the
IMF for stars $>1$\,M$_{\odot}$ can be approximated by $a=2.5\pm 0.2$
(e.g.,
\cite{pudritz2002,salpeter1955,miller1979,scalo1998,kroupa1993}). Furthermore,
\cite{motte1998} have shown that a similar slope is found between 0.5
and 3\,M$_{\odot}$ toward the young low-mass protocluster
$\rho$\,Ophiuchus. While the observations of $\rho$\,Ophiuchus already
suggested that the IMF of low-mass stars is determined at early
evolutionary stages, this was not obvious for more massive clusters
because competitive accretion and merging of intermediate-mass
protostars could establish the IMF at later stages as well (e.g.,
\cite{bonnell2004}). The new data now indicate that the upper end of
the IMF is also determined at the earliest evolutionary stages. This
supports the disk-accretion scenario for stars of all masses. However,
the observations do not rule out that coalescence might occur within
the dense centers of individual sub-ores.\\



\end{document}

%% file: beuther_refs.tex
\def\aj{AJ}%
\def\araa{ARA\&A}%
\def\apj{ApJ}%
\def\apjl{ApJ}%
\def\apjs{ApJS}%
\def\ao{Appl.~Opt.}%
\def\apss{Ap\&SS}%
\def\aap{A\&A}%
\def\aapr{A\&A~Rev.}%
\def\aaps{A\&AS}%
\def\azh{AZh}%
\def\baas{BAAS}%
\def\jrasc{JRASC}%
\def\memras{MmRAS}%
\def\mnras{MNRAS}%
\def\pra{Phys.~Rev.~A}%
\def\prb{Phys.~Rev.~B}%
\def\prc{Phys.~Rev.~C}%
\def\prd{Phys.~Rev.~D}%
\def\pre{Phys.~Rev.~E}%
\def\prl{Phys.~Rev.~Lett.}%
\def\pasp{PASP}%
\def\pasj{PASJ}%
\def\qjras{QJRAS}%
\def\skytel{S\&T}%
\def\solphys{Sol.~Phys.}%
\def\sovast{Soviet~Ast.}%
\def\ssr{Space~Sci.~Rev.}%
\def\zap{ZAp}%
\def\nat{Nature}%
\def\iaucirc{IAU~Circ.}%
\def\aplett{Astrophys.~Lett.}%
\def\apspr{Astrophys.~Space~Phys.~Res.}%
\def\bain{Bull.~Astron.~Inst.~Netherlands}%
\def\fcp{Fund.~Cosmic~Phys.}%
\def\gca{Geochim.~Cosmochim.~Acta}%
\def\grl{Geophys.~Res.~Lett.}%
\def\jcp{J.~Chem.~Phys.}%
\def\jgr{J.~Geophys.~Res.}%
\def\jqsrt{J.~Quant.~Spec.~Radiat.~Transf.}%
\def\memsai{Mem.~Soc.~Astron.~Italiana}%
\def\nphysa{Nucl.~Phys.~A}%
\def\physrep{Phys.~Rep.}%
\def\physscr{Phys.~Scr}%
\def\planss{Planet.~Space~Sci.}%
\def\procspie{Proc.~SPIE}%
\let\astap=\aap
\let\apjlett=\apjl
\let\apjsupp=\apjs
\let\applopt=\ao